\documentstyle[12pt]{article}
\begin{document}

\renewcommand{\Large}{\large}

\font\fortssbx=cmssbx10 scaled \magstep2
\hbox to \hsize{
\hskip.5in \raise.1in\hbox{\fortssbx University of Wisconsin - Madison}
\hfill\vtop{\hbox{\bf MAD/PH/746}
            \hbox{\bf SNUTP 93-7}
            \hbox{\bf YUMS 93-2}
            \hbox{\bf UMN-TH-1132/93}
            \hbox{(July 1993)}} }

\vspace{ .75in}

\begin{center}
{\large\bf Flavor Dependence of Sea Quark Structure Functions}
\\[.4in]
C.~S.~Kim,\footnotemark\
\footnotetext{Permanent address: Department of Physics, Yonsei University,
Seoul 120-749, Korea}
Sun Myong Kim,\footnotemark\
\footnotetext{Present address: Department of Physics, University of Minnesota,
Minneapolis, MN 55455}
and M.~G.~Olsson\\[.2in]
{\it Department of Physics, University of Wisconsin, Madison, WI 53706}
\end{center}

\vspace{.5in}

\begin{abstract}\normalsize
Quark masses are shown to play an important role in the nucleon sea. Our
analysis using massive QCD splitting functions demonstrates the existence of
two Feynman-$x$ sea regimes. For small $x$ the strange sea is larger while at
large $x$ the non-strange light sea is dominant. This crossover effect has been
observed in a phenomenological analysis by the CTEQ Collaboration. We also
investigate the charm component in the nucleon.
\end{abstract}

\thispagestyle{empty}

\newpage
Investigations of hadron structure functions normally assume some form of SU(3)
flavor symmetry of the sea-quark distributions. In a recent analysis of
neutrino production of unlike-sign dileptons the CCFR Collaboration~\cite{ccfr}
found the strange-sea distributions at $\left<Q^2\right>\simeq20$~GeV$^2$ to be
about one half the light-quark sea. A measure of the relative strange sea
content as a function of Feynman $x$ is given by
\begin{equation}
\kappa_s(x) = {2s(x)\over \bar u(x) + \bar d(x)} \;.
\end{equation}
If $\langle q\rangle$ is the momentum fraction carried by a quark,
$\int_0^1 x q(x) dx$, the CCFR conclusion is
$\langle s\rangle \sim {1\over2}\langle\bar u\ {\rm or}\ \bar d \rangle$ and
hence
$\langle\kappa_s\rangle = 2 \langle s \rangle / \langle\bar u + \bar d
\rangle \sim{1\over2}$.
Following the CCFR result, a number of structure function analyses have been
proposed with $\langle\kappa_s\rangle\sim{1\over2}$ at $Q=Q_0$ and
$s(x)<\bar u(x) {\rm or}\  \bar d(x)$ at all $Q$,  as shown in Fig.~1 with
dot-dashed curve for MRS-D0~\cite{mrs}, for example.

A recent global analysis by the CTEQ Collaboration~\cite{cteq} using all
relevant data finds an improved fit by a more flexible sea parameterization.
They find a crossover point for the relative sea content. For small $x$ the
strange sea is larger (and $\kappa_s>1$) while at larger $x$ the situation
reverses, as shown in Fig.~1 by the dashed curve. In addition, because of
increased  $s(x)$ in the small-$x$ region, $\left<\kappa_s\right>\simeq0.9$
instead of 0.5.

In this paper we propose a natural explanation for the seemingly unexpected
CTEQ result. Our main assertion is that quark-mass threshold effects persist to
surprisingly large $Q^2$ values, although they eventually disappear. The CTEQ
results concerning the strange sea can be qualitatively accounted for within a
straight-forward  perturbative QCD framework. We begin with a brief discussion
of our assumptions and method.

In the QCD-improved parton model, the evolution of a quark density $q(x,Q^2)$
by the Altarelli-Parisi evolution equation~\cite{altar} becomes
\begin{equation}
{dq\over d\ln Q^2} = {\alpha_s\over2\pi} \int_x^1 {dy\over y} \left[ q(y,Q^2)
P_{qq}(z) + g(y,Q^2) P_{qg}(z) \right] \,,
\end{equation}
where $z=x/y$. Since all sea quarks are generated from quark pair production by
gluons, the evolution equation for any sea quark distribution $S(x,Q^2)$ to
first order in $\alpha_s$ can be approximated as
\begin{equation}
{dS\over d\ln Q^2} \simeq {\alpha_s\over 2\pi} \int_x^1 {dy\over y} g(y,Q^2)
P_{Sg}(z) \,.
\end{equation}
And the solution to leading order for $Q>Q_0\gg\Lambda$ is
\begin{equation}
S(x,Q^2) \approx {\alpha_s(Q^2)\over2\pi} \ln{Q^2\over\Lambda^2} \int_x^1
{dy\over y} g(y,Q^2) P_{Sg}(z) \,.
\end{equation}
In the above equation $P_{Sg}$ is the Altarelli-Parisi splitting function of
gluon to sea quark $S$. For production of a massive sea--quark pair the
splitting function has the form~\cite{gluck}
\begin{eqnarray}
P_{Sg} \left(z,{m_S\over Q}\right) &=& {1\over v} \left[ {z^2+(1-z)^2\over2}
+ {m_S^2\over Q^2} \, {z(3-4z)\over 1-z}
- {16m_S^4\over Q^4}\,z^2 \right] \nonumber\\
&& {}- \left[ {2m_S^2\over Q^2}\, z(1-3z) - {8m_S^4\over Q^4}\,z^2\right]
\ln{1+v\over1-v} \,, \label{massive}
\end{eqnarray}
where $v^2= 1-{4m_S^2 z/ Q^2(1-z)}$, with $m_S=$ the mass of any sea quark $S$.
In the massless quark limit (or $Q^2\gg m_S^2$) the splitting function
(\ref{massive}) clearly reduces to the usual (flavor-independent) massless
result
\begin{equation}
P_{Sg}(z,\,m_S=0) = {z^2+(1-z)^2\over2} \,. \label{massless}
\end{equation}

The effect of a strange sea-quark threshold is most clearly displayed by
considering the ratio
\begin{equation}
\kappa_s(x,Q^2) \approx {s(x,Q^2)\over \bar u \hbox{ or } \bar d(x,Q^2)} =
{S(x, m_S=m_s, Q^2)\over S(x, m_S=m_u, Q^2)} \,. \label{ratio}
\end{equation}
The above expression has the advantage of being quite insensitive to
the details of the input gluon distribution~\cite{we} for $S$ of eq. (4).
The result is shown in Fig.~1
at $Q^2=5$~GeV$^2$ with $m_s=0.5$~GeV, $m_u=0$ (for solid curve) and $m_s=0.5$~
GeV,  $m_u=0.3$~GeV (for dotted curve). The dashed curve is the CTEQ1M
parameterization result~\cite{cteq}, and the dot-dashed is the MRS-D0
result~\cite{mrs}.  We observe the crossover effect due to the quark pair
threshold. And because of this rapid increase at small $x$, the integrated
total content of the strange sea is as large as $\bar u$ or $\bar d$ at any
$Q>Q_0$.  We find that $\left< \kappa_s (Q^2=5\,{\rm GeV}^2)\right>\simeq0.8$
integrated over $0.01<x<0.8$.  The contribution to $\left< \kappa_s \right>$
from higher than $x=0.8$ is negligible since the strange quark distribution is
much suppressed in the region.
The unusual persistence of the quark threshold at
large $x$ originates in the $1/v$ singularity of eq. (\ref{massive}).

In Fig.~2 we exhibit the sea quark threshold effect of (\ref{ratio}) for
$Q^2=20\rm~GeV^2$.  The strange to non-strange ratio
$\kappa_s$ correspond to $m_s=0.5$~GeV, $m_u=0$ (solid curve) and
$m_s=0.5$~GeV,
$m_u=0.3$~GeV (dotted curve) respectively. The dashed curve shows a charm quark
relative distribution $\kappa_c(x)$ with $m_c=1.5$~GeV and $m_u=0$. All of
these
$\kappa_{s,c}$ exhibit the crossover effect but at different Feynman $x$.
Again we can see the importance of charm content at very small $x$, which
should
be closely examined at future super colliders and deep inelastic experiments.

We have seen that due to the quark threshold effect the ratio of heavy quark to
light quark  sea components should vary considerably with Feynman~$x$.  In a
recent phenomenological data fit the CTEQ collaboration~\cite{cteq}
has observed this behavior. This variation is in
distinct contrast to previous works where $\kappa_s\sim1/2$ has been assumed.

Although we do not know the precise definition of the mass of the partons
inside the nucleon, we assume that mass differences among quarks provide the
only dynamical mechanism generating the differences of sea distribution among
quarks. We then can qualitatively reproduce the CETQ results. There still exist
many theoretically and experimentally unresolved problems related to the
understanding of the proton's structure, and we urge particle physicists to
consider seriously this problem of flavor dependence of sea distribution inside
the proton.

\section*{Acknowledgments}
We would like to thank to J.~Botts, C.~P.~Yuan and D.~Zeppenfeld for helpful
discussions, also to K. Hagiwara for critical reading of the manuscript.
C.~S.~Kim and S.~M.~Kim would like to thank the phenomenology group
at the University of Wisconsin for the warm hospitality extended to them.
This work was supported in part by the U.S.~Department of Energy under Contract
Nos.~DE-AC02-76ER00881 and DE-AC02-83ER40105, and in part by the University of
Wisconsin Research Committee with funds granted by the Wisconsin Alumni
Research Foundation. The work of C.~S.~Kim was also supported in part by the
Korean Science and Engineering Foundation, in part by Korean Ministry of
Education, in part by the Center for Theoretical Physics of Seoul National
University, and in part by a Yonsei University Faculty Research Grant.

\newpage

\section*{Figures}

\begin{enumerate}
\item
 Relative strange to non-strange components of the sea at $Q^2=5$~GeV$^2$. The
dashed curve is the CTEQ1M~\cite{cteq} phenomenological fit to a large data
set.
The crossover near $x=0.1$ is the new feature we address in this paper. The sea
quark threshold effect of Eq.~(\ref{ratio}) is shown for $m_s=0.5$~GeV,
$m_u=0$ (solid curve) and $m_s=0$, $m_u=0.3$~GeV (dotted curve).
The dot-dash curve represents the MRS-D0~\cite{mrs} parameterization evolved
from $\kappa_s=0.5$ at $Q^2=4$~GeV$^2$.

\item
Quark threshold sea enhancement $\kappa_{s,c}$ at $Q^2=20$~GeV$^2$ for charm
quark pair with $m_c=1.5$~GeV, $m_u=0$ (dashed curve), strange quark pair with
$m_s=0.5$~GeV,
$m_u=0$ (solid curve) and  $m_s=0.5$~GeV, $m_u=0.3$~GeV (dotted curve).

\end{enumerate}

\end{document}